\documentclass[twocolumn,showpacs,preprintnumbers,amsmath,amssymb]{revtex4}
\usepackage{graphicx}% Include figure files
\usepackage{dcolumn}% Align table columns on decimal point
\usepackage{bm}% bold math

\newcommand{\bemm}{\begin{multline}}
\newcommand{\enm}{\end{multline}}
\newcommand{\beq}{\begin{equation}}
\newcommand{\eeq}{\end{equation}}
\newcommand{\beqa}{\begin{eqnarray}}
\newcommand{\eeqa}{\end{eqnarray}}
\newcommand{\ba}{\begin{array}}
\newcommand{\ea}{\end{array}}

\newtheorem{teo}{Theorem}

\newcommand{\be}{\begin{equation}}

\newcommand{\ee}{\end{equation}}
\newcommand{\bt}{\begin{teo}}
\newcommand{\et}{\end{teo}}
\newcommand{\ep}{\epsilon}
\newcommand{\s}{\sigma}
\newcommand{\la}{\lambda}
\newcommand{\om}{\omega}
\newcommand{\ii}{{\rm i}}
\newcommand{\imag}{{\rm i}}

\newcommand{\bn}{B_0}
\newcommand{\bk}{B_1}
\newcommand{\an}{A_0}
\newcommand{\ak}{A_1}
\newcommand{\on}{\omega_0}
\newcommand{\ok}{\omega_1}
\newcommand{\onp}{\omega'_0}
\newcommand{\okp}{\omega'_1}
\newcommand{\si}{\sin \left(\frac {s_1}\epsilon \right)}
\newcommand{\co}{\cos \left(\frac {s_1}\epsilon \right)}

\begin{document}

%\preprint{APS/123-QED}

\title{Exact Analysis of the Adiabatic Invariants in Time-Dependent
Harmonic Oscillator}% Force line breaks with \\

\author{Marko Robnik} 
\author{Valery G. Romanovski}
\affiliation{
CAMTP - Center for Applied Mathematics and Theoretical Physics,
University of Maribor, Krekova 2, SI-2000 Maribor, Slovenia\\
}%

\date{\today}% It is always \today, today,

\begin{abstract}
The theory of adiabatic invariants has a long history and
important applications in physics but is rarely rigorous.
Here we treat exactly the general 
time-dependent 1-D harmonic oscillator,
$\ddot{q} + \omega^2(t) q=0$ which cannot be solved in general.
We follow the time-evolution of an initial ensemble of 
phase points with sharply defined energy $E_0$
and calculate rigorously the distribution of energy $E_1$ after time $T$,
and all its moments, 
especially its average value $\bar{E_1}$ and variance $\mu^2$.
Using our exact WKB-theory to all 
orders we get the exact result for the leading asymptotic behaviour
of $\mu^2$.
\end{abstract}

\maketitle

%\section{Transition map}

Adiabatic invariants, usually denoted by $I$, 
in time dependent dynamical systems (not necessarily Hamiltonian), 
are approximately conserved  during a slow process of changing system
parameters over a long typical time scale $T$. This statement
is asymptotic in the sense that the conservation is exact in
the limit $T \rightarrow \infty$, whilst for finite $T$ we 
see the deviation $\Delta I = I_{f}-I_{i}$ of final value of $I_{f}$ 
from its initial value $I_{i}$ and would like to calculate
$\Delta I$. Here we just remind that for the one-dimensional 
harmonic oscillator it is known since Ehrenfest  that
the adiabatic invariant for $T=\infty$ is $I=E/\omega$, which is
the ratio of the total energy $E=E(t)$ and the frequency of the oscillator
$\omega (t)$, both being a function of time. Of course, $2\pi I$ is
exactly the area in the phase plane $(q,p)$ enclosed by the energy contour
of constant $E$. 
A general introductory account can be found in \cite{Rob}
and references therein, especially \cite{LL,Rein}.
However, in the literature this $\Delta I$ is not even precisely defined.
As a consequence of that there is considerable confusion about
its meaning. Let us just mention the case of periodic parametric
resonance in one-dimensional harmonic oscillator where the driving 
is periodic and yet e.g. the total energy of the system can grow
indefinitely for certain system parameter values. (In this work we
give a precise meaning to these and similar statements.)
Therefore to be on rigorous side we must carefully define what we mean by
$\Delta I$. This can be done by considering an ensemble of initial
conditions at time $t=0$ just before the adiabatic process starts.
Of course, there is a vast freedom in choosing such ensembles.
In an integrable conservative Hamiltonian system the most natural 
and the most important 
choice is taking as the initial ensemble all phase points 
uniformly distributed on the initial $N$-torus, uniform w.r.t. the angle
variables. Such an ensemble has a sharply defined initial energy $E_0$.
Then we let the system evolve in time, not necessarily slowly,
and calculate the probability distribution $P(E_1)$ of the final energy $E_1$,
or of other dynamical quantities.

This is in general a difficult problem, but  in this work we confine
ourself to the one-dimensional general time-dependent harmonic oscillator,
so $N=1$, described by the Newton equation  

\be \label{Newton}
\ddot{q} + \omega^2 (t) q =0
\ee
and work out rigorously $P(E_1)$. Given the general 
dependence of the oscillator's frequency $\omega(t)$ on time $t$ the
calculation of $q(t)$ is already a very difficult unsolvable problem. 
In the sense of
mathematical physics  \eqref{Newton}
is exactly equivalent to the one-dimensional
stationary Schr\"odinger equation: the coordinate $q$ appears
instead of the probability amplitude $\psi$, time $t$ appears instead of the
coordinate $x$ and $\omega^2(t)$ plays the role of $E-V(x)$ = energy --
potential. In this paper we solve the above stated  problem for the 
general one-dimensional harmonic oscillator, but the details of
our calculations are delegated to another publication \cite{RR2005}. 
 
We begin by defining the system by giving its Hamilton function 
$H = H(q,p,t)$, whose numerical value $E(t)$ at time $t$ is precisely the total
energy of the system at time $t$, and for the one-dimensional 
harmonic oscillator this is

\be \label{Hamilton}
H=\frac {p^2}{2M}+\frac 12 M \omega^2(t)q^2,
\ee
where $q,p,M,\omega$ are the coordinate, the momentum, the mass and
the frequency of the linear oscillator, respectively. 
The dynamics is linear in $q,p$, as described by \eqref{Newton},
but nonlinear as a function of $\omega (t)$ and therefore is subject to
the nonlinear dynamical analysis. By using the index $0$ and $1$ we denote the
initial ($t=t_0$) and final $(t=t_1)$ value of the variables, 
and by $T=t_1-t_0$ we denote  the time interval of changing the parametrs of the system.

We consider the phase flow  map (we shall call it transition map)
\be
\Phi:\left( 
\begin{array}{c}
q_0\\ p_0
\end{array}
 \right)\mapsto  \left( 
\begin{array}{c}q_1\\ p_1
\end{array}
 \right).
\ee
 Because equations of motion are linear in $q$ and $p$, and since
the system is Hamiltonian,
$\Phi$ is a linear  area preserving map, that is, 
\be \label{transitionmap}
\Phi=\left(
\begin{matrix}
 a & b\\
  c & d
\end{matrix}
\right),
\ee
with  $\det(\Phi)=ad-bc=1$.
Let $E_0=H(q_0,p_0,t=t_0)$ be the initial energy and 
$E_1=H(q_1,p_1,t=t_1)$ be the final energy, that is,

\begin{multline} \label{em1}
E_1=\frac 12 \left(\frac { (cq_0+d p_0)^2}M+M \omega^2_1(a q_0+b p_0)^2\right).
\end{multline}
Introducing the new coordinates, namely the action $I=E/\omega$ and
the angle $\phi$, 
\be
 q_0=\sqrt{\frac{2 E_0}{M\omega_0^2}}\cos \phi,\ 
 p_0=\sqrt{{2M  E_0}}\sin \phi
\ee
 from \eqref{em1} we obtain
\be \label{ee1}
E_1=E_0(\alpha \cos ^2 \phi+\beta \sin^2 \phi +\gamma \sin 2\phi),
\ee
where 
\begin{multline} \label{abg}
\alpha=\frac{c^2}{M^2 w_0^2}+a^2 \frac{\omega_1^2}{\omega_0^2},\
 \beta =d^2+ \omega_1^2 M^2 b^2,\\
\gamma=\frac{c d}{M \omega_0} + ab M \frac{\omega_1^2}{\omega_0}.
\end{multline}
Given the uniform probability distribution of initial angles $\phi$
equal to $1/(2\pi)$, which defines our initial ensemble at time $t=0$,
we can now calculate the averages. Thus

\be \label{bare1}
\bar E_1=\frac 1{2\pi}\oint E_1d\phi=\frac {E_0}2 (\alpha+\beta).
\ee 
That yields 
$E_1-\bar E_1=E_0 (\delta \cos 2\phi+\gamma \sin 2 \phi)$
and 
\be \label{m2}
\mu^2 = 
\overline{(E_1-\bar E_1)^2}=\frac {E_0^2}2 \left(\delta^2+\gamma^2 \right),
\ee
where we have denoted $\delta=(\alpha-\beta)/2$. 

It follows from \eqref{abg}, \eqref{bare1}  that we can write 
\eqref{m2} also in the form
\be \label{sigma}
\mu^2 =
\overline{(E_1-\bar E_1)^2}=\frac {E_0^2}2 \left[\left( \frac{\bar E_1}{E_0} 
 \right)^2 - \frac{\om_1^2}{\om_0^2} \right].
\ee
It is straightforward to show that 
for arbitrary positive integer $m$, we have  
$\overline{(E_{1}-\bar E_{1})^{2m-1}}=0$ and

\be\label{emeven}
\overline{(E_{1}-\bar E_{1})^{2m}}=\frac {(2m -1)!!}{ m!}
\left( \overline{(E_{1}-\bar E_{1})^2}\right)^m.
\ee
Thus $2m$-th moment of $P(E_1)$ is equal to $(2m-1)!! \mu^{2m}/m!$,
and therefore, indeed, all moments of $P(E_1)$ are uniquely determined 
by the first moment $\bar{E_1}$.

Expression \eqref{sigma} is positive definite by definition and this
leads to the first interesting conclusion: In full generality 
(no restrictions on the function $\omega (t)$!) we have always
$\bar{E_1} \ge E_0\omega_1/\omega_0$ and therefore the final value
of the adiabatic invariant $I_1=\bar{E_1}/\omega_1$ is always greater
or equal to the initial value $I_0=E_0/\omega_0$. In other words,
the value of the adiabatic invariant never decreases, which is
a kind of irreversibility statement. Moreover, it is constant only
for infinitely slow processes $T=\infty$, which is an ideal adiabatic 
process, i.e. $\mu=0$.
For periodic processes $\omega_1=\omega_0$ we see that 
always $\bar{E_1} \ge E_0$, so the mean energy never decreases.
The other extreme to $T=\infty$ is the instantaneous ($T=0$) 
jump where $\omega_0$
switches to $\omega_1$ discontinuously, whilst $q$ and $p$ remain
continuous, and this results in $a=d=1$ and $b=c=0$, and then we find

\be \label{jump}
\bar{E_1} = \frac{E_0}{2} (\frac{\omega_1^2}{\omega_0^2} + 1),\;\;\;\;\;
\mu^2 = \frac{E_0^2}{8} \left[ \frac{\omega_1^2}{\omega_0^2} -1\right]^2.
\ee
Below we shall treat the special case with $\omega_1^2 = 2\omega_0^2$,
and thus will find $\mu^2/E_0^2 = 1/8 = 0.125$.

Our general study now focuses on the calculation of the transition
map \eqref{transitionmap}, namely its matrix elements $a,b,c,d$.
Starting from the Hamilton function \eqref{Hamilton}  and its Newton equation
\eqref{Newton}
we consider two  linearly  independent solutions $\psi_1(t)$ 
and $\psi_2(t)$ and introduce the matrix

\be \label{phig}
\Psi(t)=\left(
\begin{matrix}
 \psi_1(t) & \psi_2(t)\\
  M\dot{\psi}_1(t) & M\dot{\psi}_2(t)
\end{matrix}
\right).
\ee
Consider  a solution $\hat q(t)$ of \eqref{Newton}
such that
\be \label{qp0}
\hat q(t_0)=q_0,\quad \dot { \hat q}(t_0)=p_0/M.
\ee
Because $\psi_1$ and $\psi_2$ are linearly independent, we can look for  
 $\hat q(t)$ in the form
\be
\hat q(t)=A \psi_1(t) + B \psi_2(t). 
\ee
Then $A$ and $B$ are determined by 
\be \label{AB}
\left( 
\begin{array}{c}
A\\ B
\end{array}
 \right)= \Psi^{-1}(t_0) \left( 
\begin{array}{c}q_0\\ p_0
\end{array}
 \right).
\ee
Let  $q_1=\hat q(t_1),\quad  p_1=M \dot {\hat q}(t_1)$.
Then from  (\ref{qp0})--(\ref{AB}) we  see that
\be \label{p1q1}
\left( 
\begin{array}{c}
q_1\\ p_1
\end{array}
 \right)= \Psi(t_1)  \Psi^{-1}(t_0) \left( 
\begin{array}{c}q_0\\ p_0
\end{array}
 \right).
\ee
We recognize the matrix on the right-hand side of \eqref{p1q1} 
as the transition map $\Phi$, that is,
\be \label{phiT}
\Phi=\left(
\begin{matrix}
 a & b\\
  c & d
\end{matrix}
\right)= \Psi(t_1)  \Psi^{-1}(t_0).
\ee
Due to lack of space we mention only one specific model, namely
the linear model defined by the piecewise linear $\omega^2(t)$
function

\be
\omega^2(t)=\left\{ 
\begin{array}{l}
\omega_0^2 \quad \quad \quad \quad \quad   \quad{\ \,  \rm if }\ \ \  t\le 0\\
 \omega_0^2+ \frac{(\omega_1^2-\omega_0^2)}T\, t \quad \  {\rm if}\ \ \  0 <t < T\\
\omega_1^2 \quad \quad  \quad \quad \quad \quad  {\ \,  \rm if  }\ \ \  t\ge T
\end{array}. \right.
\ee
In this case the equation \eqref{Newton}
can be solved exactly in 
terms of the Airy functions, and the formalism explained above leads
in a straightforward  but lengthy manner to the final exact result
for $\bar{E_1}$ and consequently for $\mu^2$ etc. It is too complex
to be shown here. The special case $\omega_0^2=1$ and 
$\omega_1^2= 2$ has been checked very carefully, also numerically,
and $\mu^2$ goes correctly from $1/8$ at $T=0$ to zero as 
$T \rightarrow \infty$, in a typical oscillatory way.
 Using the well known asymptotic expressions
for the Airy functions we find the leading asymptotic approximation

\be \label{m1ap1}
\frac{\mu^2}{E_0^2}  = \frac{\overline{(E_1-\bar E_1)^2}}{E_0^2}
\approx 
\frac {\ep^2}{128}  \left( 9 - 4\,{\sqrt{2}}\,\cos (\frac{4 - 8\,{\sqrt{2}}}{3\, {\ep}})  \right),
\ee
where we introduce the adiabatic parameter  $\ep=1/T$
which is assumed small. Please observe the oscillatory approach to zero
as $\epsilon \rightarrow 0$, which in the mean goes to zero quadratically
as $\epsilon^2$.

Returning to the general case 
we now mention that the final energy distribution function written down as

\be \label{Edistribution}
P(E_1)=\frac 1{2\pi} \sum_{j=1}^4\left| \frac{d\phi}{dE_1} 
\right|_{\phi=\phi_j(E_1)}
\ee
cannot be calculated analytically in a closed form in any useful way, 
because it boils down to finding the roots of a quartic polynomial, so
we do not try to do that here, although numerically it  shows
interesting aspects. It has a finite interval as its support, between
the lower limit $E_{min}$ and the upper limit $E_{max}$, and at
both values it has an integrable singularity of the type $1/\sqrt{x}$.
 In between
for every value of $E_1 = const = E_1(\phi)$, this equation has four solutions,
namely $\phi_1,\phi_2,\phi_3,\phi_4$, and thus we have to sum up
all four contributions in the general formula \eqref{Edistribution}.
On the other hand, as we have seen, 
the moments of this interesting distribution function
can be calculated exactly to all orders.

We proceed with the calculation of the transition map $\Phi$ in
the general case, and because \eqref{Newton}
is generally not solvable,
we have ultimately to resort to some approximations. Since the adiabatic
limit $\ep \rightarrow 0$ is the asymptotic regime that we would
like to understand, the application of the rigorous WKB theory
(up to all orders) is most convenient, 
and usually it turns out that the leading
asymptotic terms are well described by just the leading WKB terms.

We introduce re-scaled and dimensionless time  $\lambda$
\be
\lambda= \epsilon t,  \;\;\;\;\; \ep = 1/T
\ee
so that \eqref{Newton} is transformed to the equation

%\footnote{ Calculations of $\mu_k$ at the end of the manuscript   
%are for the equation  
%$
% [- {\epsilon ^2} \frac {{\rm d}^2}{{\rm  d} x^2}+V(x)]\psi (x)=E\psi(x).
%$
% It is necessary to pay attention to the sign.}

\be \label{sch}
\ep^2 q''(\la)+\omega^2(\lambda)q(\la)=0.
\ee
Let $q_+(\lambda)$ and $q_-(\lambda)$  
be two linearly  independent solutions of 
\eqref{sch}. Then the matrix  \eqref{phig} takes the form 

\be \label{phig1}
\Psi_\la=\left(
\begin{matrix}
 q_+(\la) & q_-(\la)\\
\ep M q'_+(\la) & \ep M  q'_-(\la)
\end{matrix}
\right)
\ee
and taking into account that $\la_0=\ep t_0, \la_1=\ep t_1 $ we obtain for the 
matrix \eqref{phiT} the expression

\be 
\Phi=\left(
\begin{matrix}
 a & b\\
  c & d
\end{matrix}
\right)= \Psi_\la(\la_1)  \Psi_\la^{-1}(\la_0).
\ee
We now use the WKB method in order to obtain 
the coefficients $a, b, c, d$ of the matrix $\Phi$.
To do so,  we look for solution of \eqref{sch} in the form
\be
q(\la) = w \exp \left\{\frac 1\epsilon \sigma (\la)\right\}
\ee
where  $\sigma (\la)$ is a complex function that satisfies the
differential equation 
\be \label{w1}
(\sigma '(\la))^2+ \epsilon  \sigma '' (\la)=-\om^2(\la)
\ee
and $w$ is some constant with dimension of length.
The WKB expansion for the phase is
\be \label{w2}
\sigma(\la)=\sum_{k=0}^{\infty}
  \epsilon^k \sigma_k (\la).
\ee
Substituting (\ref{w2}) into (\ref{w1}) and comparing
like powers of  $\epsilon$ gives the recursion relation
\be \label{w3}
\s_0'^2 = -\omega^2(\la),\ \ \ \ \ \s'_{n}=-\frac 1{2\s'_0}
(  \sum_{k=1}^{n-1} \s' _k \s ' _{n-k}+\s_{n-1}'').
\ee
Here we apply our WKB notation and formalism \cite{RR2000}
and we can choose
$\s'_{0,+} (\la)=\ii \om(\la)\quad {\rm  or} \quad  \s'_{0,-} (\la)
=-\ii \om(\la)$.
That  results in two linearly independent solutions of \eqref{sch}  
given by the WKB  expansions with the coefficients 
\bemm
\s_{0,\pm}(\la)=\pm \ii  \int _{{\lambda_0}}^{\la} \om (x) dx, 
\ \s_{1,\pm}(\la)=
- \frac 12 \log \frac{\om(\la)}{\om(\la_0)}, \\ \s_{2,\pm}= \pm
\frac{\imag }{8}\,\int _{{\lambda_0}}^{\la}
    \frac{3\,{\om'(x)}^2 - 2\,\om(x)\,\om''(x)}
      {{\om(x)}^3}\,dx, \ \dots
\end{multline}
Since $\om (\la)$ is a real function we deduce from \eqref{w3} 
that all functions $\s'_{2k+1}$ are real and all   functions $\s'_{2k}$
are pure imaginary
and 
$\s'_{2k,+}=- \s'_{2k,-}, \quad \s'_{2k+1,+}= \s'_{2k+1,-}$
where $k=0,1,2,\dots$, and thus we have 
$\s'_+=A(\la)+\ii B(\la), \quad \s'_-=A(\la)-\ii B(\la)$
where 
$A(\la) = \sum_{k=0}^\infty \ep^{2k+1}\s'_{2k+1}(\la), \quad 
B(\la) = -\ii \sum_{k=0}^\infty \ep^{2k}\s'_{2k,+}(\la)$.
 Integration of the above equations 
yields
\be
\s_+=r(\la)+\ii s(\la), \quad \s_-=r(\la)-\ii s(\la),
\ee
where 
$r(\la)=\int_{\la_0}^\la A(x)\, dx, \quad  s(\la)=\int_{\la_0}^\la B(x)\, dx$.
Below we shall denote $s_1=s(\la_1)$.

Using this notation we find that the elements of the transition 
matrix $\Phi$ have the following form, after taking into account
that $\det (\Phi) = ab-cd=1$,

\be
\begin{aligned} \label{abcd}
a = & -\frac{1}{\sqrt{\bn\bk}} \left[\an \si -\bn \co \right],\\
b = &  \frac{1}{M \sqrt{\bn\bk}}\si,\\
c = & - \frac{M}{\sqrt{\bn\bk}  }\left[ \left (\an \ak +\bn \bk \right)\si \right. \\
& \left. 
+ \left( \an \bk - \ak \bn   \right)\co  \right],\\
d = &  \frac{1}{\sqrt{\bn\bk}  }\left[ \ak \si +\bk \co  \right]. 
\end{aligned}
\ee
This is so far exact result, based on the WKB expansion technique.
What we are mostly interested in is the asymptotic behaviour  of
$\mu^2$ when $\ep$ is small and tends to zero. All other aspects
and technical details will be published in a separate paper \cite{RR2005}.

Let us consider the first order WKB approximation,
that is,
\be
A(\la)\approx \ep \s'_{1,+}(\la), \quad B (\la) \approx \frac{\s'_{0,+}(\la)}\ii=\om(\la).
\ee  
We find for the variance \eqref{sigma}
\begin{multline}
 \frac{\mu^2}{E_0^2}  =  \ep^2 \left(
 \frac {\ok^2 {\onp}^2 }{8 \on^6}  +   \frac{{\okp}^2}{8 \on^2 \ok^2}   -
\right. \\  \left.
\frac{ \onp
 \okp }{4 \on^4  } \cos\left( \frac 2\ep  \int_{\la_0}^{\la_1} \om(x) \,dx 
 \right)\right)+O(\ep^3).
\end{multline}
Substituting into the last formula $\om(\la)= \sqrt{1+\la} $
we obtain exactly the approximation \eqref{m1ap1}.

Suppose now  that all derivatives at $\la_0$ and $\la_1$ vanish
up to order $(n-1)$, i.e.
$\om'(\la_0)=\om'(\la_1)=\dots = \om^{(n-1)}(\la_0)=\om^{(n-1)}(\la_1)=0$,
and $\om^{(n)}(\la_0)\om^{(n)}(\la_1)\ne 0$.
Then
$
\s'_1(\la_0)=\s'_1(\la_1)=\dots = \s'_{n-1}(\la_0)=\s'_{n-1}(\la_1)=0,\
\s'_{n}(\la_0)\s_{n}(\la_1)\ne 0.
$
%\end{multline}

Hence, in the case $n=2k-1$ we can assume 

\be \begin{aligned}
A (\la) =& \ep^{2k-1} \s'_{2k-1,+}(\la) + h.o.t. \\  B (\la) 
=& \om(\la)-\ii \ep^{2k}  { \s'_{2k,+}(\la)} + h.o.t.
\end{aligned}
\ee 
and obtain 
\begin{multline} \label{sim1}
  \frac {\mu^2}{E_0^2}=  
\ep^{4k-2}\left(
\frac {  \s'_{2k-1,+}(\la_1) ^2}{2 \on^2}+\frac{\ok^2  \s'_{2k-1,+}(\la_0) ^2}{2\on^4}-\right. \\ \left.  \frac{\ok  \s'_{2k-1,+}(\la_0)  \s'_{2k-1,+}(\la_1) }{\on^3}\cos\left( \frac {2 s_1}\ep\right)\right)\\ +O(\ep^{4k-1}).
\end{multline}

In the case when $n=2 k$ we can suppose  
\be 
\begin{aligned}
 A(\la)= & \ep^{2k+1} \s'_{2k+1,+}(\la)+h.o.t. \\ 
 B (\la) = &  \om(\la)-\ii \ep^{2k}  { \s'_{2k,+}(\la)}+h.o.t.
\end{aligned}
\ee
Then, similarly as above,
we obtain
\begin{multline}
  \frac {\mu^2}{E_0^2}=  - 
\ep^{4k}\left(
\frac {  \s'_{2k,+}(\la_1) ^2}{2 \on^2}+\frac{\ok^2  \s'_{2k,+}(\la_0) ^2}{2\on^4}- \right. \\ \left.  \frac{\ok  \s'_{2k,+}(\la_0)  \s'_{2k,+}(\la_1) }{\on^3}\cos\left( \frac {2s_1}\ep\right)\right) +O(\ep^{4k+1}).
\end{multline}

From this we can conclude that if $\omega (t)$ is of class ${\cal C}^{m}$
(having $m$ continuous derivatives, $m=n-1$) 
$\mu^2$ goes to zero oscillating but in the mean as 
$\propto \ep^{2n}=\ep^{2(m+1)}$.
If $m=\infty$ (analytic functions) according to  Landau  and Lifshitz 
\cite{LL}
the decay to zero is oscillating and on the average is
exponential $\propto \exp (-const/\ep)$.

\begin{acknowledgments}
This   work was supported   by the Ministry of
Higher  Education, Science and Technology 
of the Republic of Slovenia,  Nova Kreditna Banka Maribor and 
 Telekom Slovenije.
\end{acknowledgments}

\bibliography{adiabaticprl1}% Produces the bibliography via BibTeX.

\end{document}